\documentclass[10pt,conference]{IEEEtran}
\IEEEoverridecommandlockouts
\usepackage{cite}
\usepackage{amsmath,amssymb,amsfonts}
\usepackage{algorithmic}
\usepackage{graphicx}
\usepackage{textcomp}
\usepackage{xcolor}
\usepackage{listings}
\usepackage{subcaption}
\usepackage[scaled=0.95]{inconsolata} % use a better font for listings

\def\BibTeX{{\rm B\kern-.05em{\sc i\kern-.025em b}\kern-.08em
    T\kern-.1667em\lower.7ex\hbox{E}\kern-.125emX}}

\begin{document}

\author{\IEEEauthorblockN{Sungmin Kang}
\IEEEauthorblockA{
KAIST\\
sungmin.kang@kaist.ac.kr}
\and
\IEEEauthorblockN{Shin Yoo}
\IEEEauthorblockA{
KAIST\\
shin.yoo@kaist.ac.kr}
}

\title{Towards Objective-Tailored Genetic Improvement Through Large Language Models}

\maketitle

\begin{abstract}
While Genetic Improvement (GI) is a useful paradigm to improve functional and nonfunctional aspects of software, existing techniques tended to use the same set of mutation operators for differing objectives, due to the difficulty of writing custom mutation operators. In this work, we suggest that Large Language Models (LLMs) can be used to generate objective-tailored mutants, expanding the possibilities of software optimizations that GI can perform. We further argue that LLMs and the GI process can benefit from the strengths of one another, and present a simple example demonstrating that LLMs can both improve the effectiveness of the GI optimization process, while also benefiting from the evaluation steps of GI. As a result, we believe that the combination of LLMs and GI has the capability to significantly aid developers in optimizing their software.
\end{abstract}


\section{Introduction}

Software is complex, and as a result it can take significant manual effort to optimize software to better meet requirements such as runtime or memory usage. To reduce the amount of developer time that must be spent on optimizing software, the field of Genetic Improvement (GI)~\cite{Petke:2017aa} has strived to use the principles of stochastic optimization to automatically optimize software. 
Specifically, one defines a fitness function (e.g., the execution time of code) and genetic operators that change and combine source code in ways that are expected to help; at each `generation' the best solutions according to the fitness function are selected and further modified. This process is repeated until some termination criterion is met, and the optimized results are presented to a developer. Such a paradigm has led to significant successes, such as automated program specialization~\cite{Petke2013bz,Petke:2014wj}, energy consumption optimization~\cite{Bruce2015pr} and continuous automated program repair~\cite{Haraldsson2017qy}. 

Despite these successes, the traditional formulation has a major limitation in its use of genetic operators. In the traditional formulation of GI, 
the code under improvement is to be randomly modified then to be compiled and evaluated for the given objective in an automated pipeline. Any 
change that is not compilable will block the pipeline, significantly reducing the overall efficiency. Consequently, much existing 
work~\cite{Petke2013bz,Petke:2014wj,Haraldsson2017qy} relies on the same genetic operators first investigated by GenProg~\cite{Weimer:2009fk}, i.e., inserting code borrowed somewhere else in the same codebase, deleting code, or swapping two existing code elements. While this approach improves the probability of successful compilation of the randomly modified code, it also has two limitations. First, the change is not customized for the given objective, leaving us simply to hope for the existence of some ingredients that can contribute to the given objective. Second, this approach is known to also generate \emph{bloated} results, requiring additional post-processing to minimize the results~\cite{Langdon:2012fk}.

Meanwhile, large language models (LLMs) are showing impressive performance in natural language processing and software engineering tasks~\cite{brown2020language,Chen2021ec}; one can ask an LLM to change code in a specific manner using natural language (as we later show with an example), resulting in objective-specific genetic operators. Furthermore, an LLM trained on software defines a distribution over code, and as a result generates code with high naturalness~\cite{Hindle:2012kq}, improving the probability of successful compilation as well as that of the results being accepted in practice.

This is not to say LLMs are a panacea -- while LLMs generate plausible improvements, these improvements are not always correct, and can suffer from myriad mistakes~\cite{Sarkar2022PwAI}. Perhaps most critically, while GI-generated solutions that have a better fitness score than the original are verified solutions that yield better results on what the fitness score measures, LLM-generated solutions are not guaranteed to improve the software, at least not without a verification step.

This is why we believe LLMs and the GI process have significant synergy: working together, each component would augment the strengths and complement the weaknesses of the other. That is, LLM+GI would allow objective-specific modifications to be added to the GI candidate pool via the flexibility of LLMs; LLM+GI can ensure that the desired objective is met by the rigor of the evolutionary cycle.

In the next section, we provide an example of an LLM to improve inefficient Python implementations of Fibonacci number calculation, to demonstrate the significant potential of LLMs when incorporated in the GI process.

\section{Demonstration}

As a simple demonstration of LLMs, we choose two objectives from the GI Call for Papers: ``improv[ing] efficiency'' (which we understand as execution time) and ``decreas[ing] memory consumption''. We manually write inefficient implementations for calculating the Fibonacci numbers on each aspect, as shown in Figure~\ref{fig:ineff_code}. Figure~\ref{fig:ineff_code}(a) leads to an $O(\phi^n)$ computational complexity (where $\phi=\frac{1+\sqrt{5}}{2}$), whereas Fibonacci numbers can be calculated in $O(n)$ time; Figure~\ref{fig:ineff_code}(b) leads to an $O(n)$ memory complexity, whereas Fibonacci numbers can be calculated using $O(1)$ memory. 

\begin{figure}[ht!]
    \centering
    \begin{subfigure}[t]{0.23\textwidth}
        \caption{Time-inefficient code.}
        \begin{lstlisting}[basicstyle=\footnotesize\ttfamily,
            columns=flexible,
            breaklines=true,
            language=python]
def fibonacci(n):
  if n == 1 or n == 2:
    return 1
  else:
    return fibonacci(n-1)+fibonacci(n-2)
        \end{lstlisting}
    \end{subfigure}
    \begin{subfigure}[t]{0.23\textwidth}
        \caption{Memory-inefficient code.}
        \begin{lstlisting}[basicstyle=\footnotesize\ttfamily,
            columns=flexible,
            breaklines=true,
            language=python]
def fibonacci(n):
  l = [1, 1]
  while len(l) < n:
    l.append(l[-1]+l[-2])
  return l[-1]
        \end{lstlisting}
        
    \end{subfigure}
    \caption{Inefficient implementations of the Fibonacci numbers.}
    \label{fig:ineff_code}
\end{figure}

We prompt the LLM to make these implementations more efficient by specifying the objectives in natural language. Specifically, we add a comment above the original code saying `original, [time/memory]-inefficient code', then add a comment below the code saying `fixed, [time/memory]-efficient code', along with a function name indicating the property of the implementation (see Figure~\ref{fig:example_prompt} for an example). This prompt is then submitted to the LLM to get modified code. We used the \texttt{code-davinci-002} model from OpenAI, under the default parameter settings.

\begin{figure}[ht!]
  \centering
  \begin{subfigure}[t]{0.48\textwidth}
      \begin{lstlisting}[basicstyle=\footnotesize\ttfamily,
          columns=flexible,
          breaklines=true,
          language=python]
# original, time-inefficient code
def fibonacci(n):
  (original code)

# fixed, time-efficient code
def fibonacci_fast(n):
      \end{lstlisting}
  \end{subfigure}
  \caption{Example prompt for time optimization.}
  \label{fig:example_prompt}
\end{figure}

The results are presented in Figure~\ref{fig:codex_fixed}: each version is more efficient than the original in the way intended. In Fig.~\ref{fig:codex_fixed}(a), note that the code with exponential time complexity was replaced with an $O(n)$ time complexity algorithm that is fast. Meanwhile, Fig.~\ref{fig:codex_fixed}(b) shows a memory-efficient implementation that only uses two variables to calculate Fibonacci numbers.

\begin{figure}[h!]
  \centering
  \begin{subfigure}[t]{0.23\textwidth}
      \caption{Time-efficient code.}
      \begin{lstlisting}[basicstyle=\footnotesize\ttfamily,
          columns=flexible,
          breaklines=true,
          language=python,]
def fibonacci_fast(n):
if n == 1 or n == 2:
  return 1
else:
  a = 1
  b = 1
  for i in range(3,n+1):
    c = a + b
    a = b
    b = c
  return c
      \end{lstlisting}
      
  \end{subfigure}
  \begin{subfigure}[t]{0.23\textwidth}
      \caption{Memory-efficient code.}
      \begin{lstlisting}[basicstyle=\footnotesize\ttfamily,
          columns=flexible,
          breaklines=true,
          language=python,]
def fibonacci_efficient(n):
a, b = 1, 1
for i in range(n-1):
  a, b = b, a+b
return a
    
      \end{lstlisting}
      
  \end{subfigure}
  \caption{LLM-generated efficient Fibonacci implementations.}
  \label{fig:codex_fixed}
\end{figure}

In our examples, all implementations generated by the LLM are correct, as the Fibonacci calculation problem is well-known and doubtless part of its training data. Nonetheless, LLMs are prone to generating incorrect outputs as Sarkar et al.~\cite{Sarkar2022PwAI} note, especially when the code to optimize becomes complex. Further, the verification step adds value to developers: Winter et al.~\cite{Winter2022td} find that developers are likely to be persuaded by objective metrics. As a result, we believe that LLM output can become significantly more valuable when combined with the GI process.

\section{Conclusion}

In this work, we argue that there is significant synergy between large language models and genetic improvement. To this end, we provide an example demonstrating that LLMs can act as effective mutators of source code given a natural language description, while also suggesting that LLM output would benefit from the rigorous evaluation of GI as well. Such results demonstrate the possibility of using LLMs to significantly reduce developer effort when optimizing software, coming closer to the overall goal of automatic software improvement; as a result, we believe there is a bright future for combined techniques utilizing LLMs and GI.

\section*{Acknowledgment}
This work was supported by the National Research Foundation of Korea (NRF) Grant
(NRF-2020R1A2C1013629).

\bibliographystyle{IEEEtran}
\bibliography{newref}

\end{document}